1# Rapid Satellite-to-Site Visibility Determination Based on Self-Adaptive Interpolation Technique

HAN Chao, GAO Xiaojie, SUN Xiucong [*]
*School of Astronautics, Beihang University, Beijing 100191, China***Abstract**

Rapid satellite-to-site visibility determination is of great significance to coverage analysis of satellite constellations as well as onboard mission planning of autonomous spacecraft. This paper presents a novel self-adaptive Hermite interpolation technique for rapid satellite-to-site visibility determination. Piecewise cubic curves are utilized to approximate the waveform of the visibility function versus time. The fourth-order derivative is used to control the approximation error and to optimize the time step for interpolation. The rise and set times are analytically obtained from the roots of cubic polynomials. To further increase the computational speed, an interval shrinking strategy is adopted via investigating the geometric relationship between the ground viewing cone and the orbit trajectory. Simulation results show a 98% decrease in computation time over the brute force method. The method is suitable for all orbital types and analytical orbit propagators.

*Keywords:* satellite-to-site visibility; rapid determination; self-adaptive Hermite interpolation; viewing cone[*] Corresponding author. (e-mail: sunxiucong@gmail.com)

## 1. Introduction

The satellite-to-site visibility problem, which refers to the determination of opportunities for a satellite to observe or communicate with an object on the Earth's surface, plays an important role in coverage analysis of satellite orbits and constellations [1,2]. For example, in order to assess coverage performance of an Earth observation constellation, it is first necessary to generate the visibility periods of each ground target for all the member satellites. In addition, with the increasing capability of onboard computers, autonomous mission planning is projected for future generation of spacecraft systems. Onboard prediction of satellite-to-site visibility will be required for intelligent Earth imaging tasks as well as for data communication scheduling [3,4].

Satellite-to-site visibility periods are typically determined by the conventional brute force method, which is to let the satellite run through its trajectory and to check whether it can access the site at each instant. The disadvantage of this method is that orbital positions are calculated hundreds of times per orbital period and thus the execution time is tremendous, especially when perturbation effects are considered. Even though the computation load is tolerable for ground facilities, it is undesirable for onboard real-time mission planning. Thus, the development of rapid algorithms for visibility computation is crucial to both ground-based constellation design and onboard spacecraft autonomy.

Fast algorithms have been proposed by many researchers in order to reduce the computation cost of satellite-to-site visibility determination. A closed-form solution is given by Escobal [5] for two-body motion by introducing a single transcendent equation as a function of eccentric anomaly. The equation is solved only once per orbital period. Mai and Palmer [6] presented a coarse-to-fine strategy via checking the closest satellite ascending pass over the target latitude. Analytical perturbations are incorporated to improve visibility determination accuracy. However, the method is not applicable for low-inclined orbits. The idea of using closest pass of satellite is also utilized in [7], but the satellite ground trace is approximated by a great-circle arc. Thus the method is only valid for circular orbits. Apart from these geometric approaches, numerical methods based on visibility function approximation have been developed. In [8], the Fourier series yields a good representation of visibility function for low-eccentricity orbits. Alfano et al. [9] further used the parabolic blending technique to construct the waveform of visibility function with a fixed time step to suit all orbital types. A 95% decrease in computation time over the brute force method was achieved. More recently, Sun et al. [10] improved Alfano's method and developed an adaptive Hermite interpolation technique which decreases the time step of curve fitting to guarantee interpolation accuracy at fast-changing parts of the waveform. The adaptive Hermite interpolation technique has also been employed for rapid Geometric Dilution of Precision (GDOP) analysis of navigation constellations [11]. However, theoretical proof of the method is not given and the time step of curve fitting is not adaptively increased at smooth parts of the waveform. Thus its computational efficiency is slightly inferior to that of the parabolic blending technique.

In this paper, a self-adaptive Hermite interpolation technique in the strict sense is presented for curve fitting of the satellite-to-site visibility function. The fourth-order derivative is utilized to control the approximation error and to adaptively adjust the time step of interpolation. In addition, the geometric relationship between the ground viewing cone and the satellite's space trajectory is used to identify time intervals during which the satellite will never see the ground station. In Section 2, we introduce the mathematical model of the satellite-to-site visibility problem. The self-adaptive interpolation technique is described and the algorithm design for rapid visibility determination is presented. In Section 3, we introduce the interval shrinking strategy based on geometric analysis of the ground viewing cone. Section 4 presents simulation results. Conclusion is drawn in Section 5.

## 2. Rapid determination of satellite-to-site visibility

We first introduce the mathematical model of the satellite-to-site visibility problem and generalize it as a multiple hump function rooting problem. Then we present the basic principle of the self-adaptive cubic interpolation method for rapid root finding. The detailed algorithm for the satellite-to-site visibility calculation is given thereafter.

## 2.1. Mathematical model

A satellite cannot see a ground target until it rises above a minimum elevation angle. Let $\theta$ denote the elevation angle of the satellite at the current position, and $\theta_0$ denote the predefined minimum elevation. As illustrated in Fig. 1, $\pi/2 - \theta$ equals the angle between the position vector of the ground target and the relative position vector from the ground target to the satellite. The visibility criterion is thus given as follows

$$\frac{\Delta \boldsymbol{r} \cdot \boldsymbol{r}_{site}^0}{\|\Delta \boldsymbol{r}\|} \geq \sin \theta_0 \qquad (1)$$

where $\Delta \boldsymbol{r} = \boldsymbol{r}_{sat} - \boldsymbol{r}_{site}$, $\boldsymbol{r}_{site}^0 = \boldsymbol{r}_{site}/\|\boldsymbol{r}_{site}\|$, $\boldsymbol{r}_{sat}$ and $\boldsymbol{r}_{site}$ are the position vectors of the satellite and the ground target, respectively, and $\|\cdot\|$ denotes the magnitude of a vector. The value of the term on the left side of Eq. (1) varies with time and is defined as the visibility function, denoted by $V(t)$, whereas the term on the right side is a constant and is defined as the visibility threshold, denoted by $\lambda$.

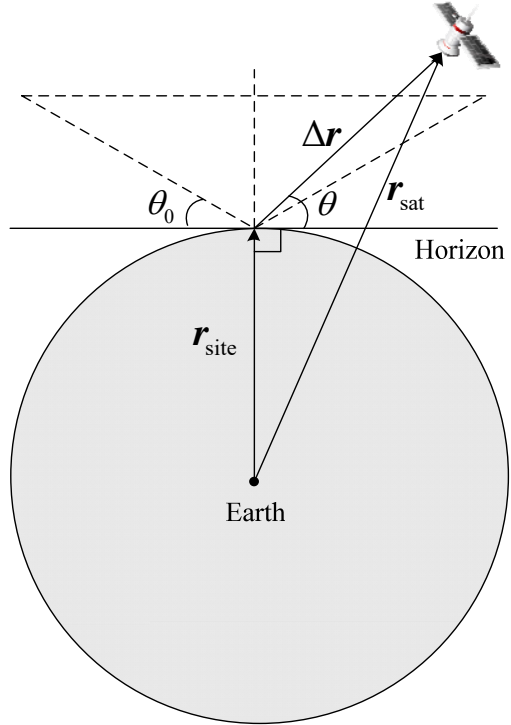

**Figure 1**. Geometric illustration of the satellite-to-site visibility.

The rise and set times are the solutions of the following nonlinear equation

$$V(t) - \lambda = 0 \qquad (2)$$

Assume a circular orbit with an altitude of 1000 km and a ground target located at (120° E, 40° N). Figure 2 shows the time-varying behavior of $V(t)$ as well as a visibility threshold corresponding to a minimum elevation of 10°. The satellite sees the ground target only when the visibility function value is above the threshold line. As depicted in Fig. 2, the visibility function is a multiple hump function, which is erratic and has no fixed periods for non-regressive orbits. There are in general no analytical expressions for the roots of such a system.

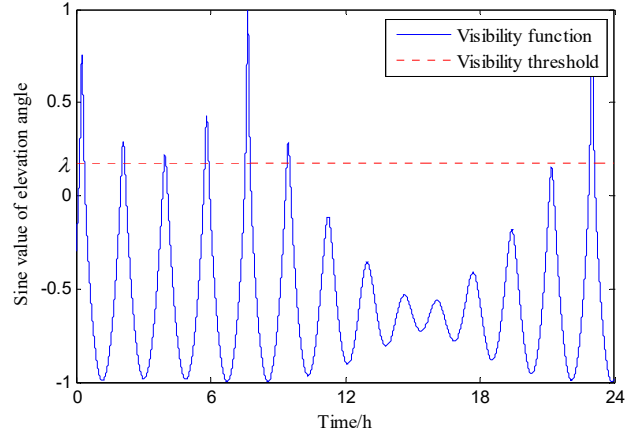

**Figure 2**. Satellite-to-site visibility function as a multiple hump function.

The piecewise cubic lines have been used to approximate the visibility function in [10,11]. Cubic curves normally have two turning points and could fit the visibility function well near the peaks. In addition, cubic curves have analytical expressions of roots.

Consider a visibility function $V(t)$ with $t \in [t_1, t_2]$. Let $\dot{V}(t)$ denote its first-order derivative function. Given an interval partition $\Delta = \{t_1 = \xi_0 < \xi_1 \cdots < \xi_n = t_2\}$, cubic Hermite interpolation can be used to create cubic polynomials on each subinterval to approximate $V(t)$ using the visibility function values as well as the first-order derivatives at two bounds

$$C_i(t) = \frac{3h_i(t-\xi_{i-1})^2 - 2(t-\xi_{i-1})^3}{h_i^3} V(\xi_i) + \frac{h_i^3 - 3h_i(t-\xi_{i-1})^2 + 2(t-\xi_{i-1})^3}{h_i^3} V(\xi_{i-1}) \\ + \frac{(t-\xi_{i-1})^2(t-\xi_i)}{h_i^2} \dot{V}(\xi_i) + \frac{(t-\xi_{i-1})(t-\xi_i)^2}{h_i^2} \dot{V}(\xi_{i-1}) \quad (3)$$

where $C_i(t)$ is the cubic function approximating $V(t)$ on the $i$th subinterval $[\xi_{i-1}, \xi_i]$, and $h_i = \xi_i - \xi_{i-1}$ is the time step.

The overall approximation accuracy is directly related to the choice of interpolation points, $\xi_0, \xi_1, \cdots, \xi_n$. Increase in the number of interpolation points leads to improved accuracy. However, the computational speed is meanwhile decreased. Within this study, the compromise between accuracy and efficiency is achieved using a self-adaptive step searching approach, where the fourth-order derivative is utilized to optimize the choice of interpolation points.

According to [12], the approximation error can be described by

$$R(t) = V(t) - C_i(t) = \frac{V^{(4)}(\eta)}{4!} \left[(t-\xi_{i-1})(t-\xi_i)\right]^2, \quad t, \eta \in [\xi_{i-1}, \xi_i] \quad (4)$$

where $V^{(4)}(t)$ is the fourth-order derivative of $V(t)$ and can be obtained using higher-order Hermite interpolation. Here we first construct a quintic polynomial $P(t)$ using the visibility function values and the first-order derivatives at three points $\xi_{i-1}$, $\xi_{i-1} + h_i/2$, and $\xi_i$, and then utilize its fourth-order derivative to approximate $V^{(4)}(t)$

$$V^{(4)}(t) \approx P^{(4)} = 120 a_5 t + 24 a_4 \quad (5)$$

where $a_5$ and $a_4$ are the coefficients of the fifth and fourth order terms of $P(t)$ and are given by

$$a_5 = \frac{24}{h_i^5}\left[V(\xi_{i-1}) - V(\xi_i)\right] \\ + \frac{4}{h_i^4}\left[\dot{V}(\xi_{i-1}) + 4\dot{V}\left(\xi_{i-1} + \frac{1}{2}h_i\right) + \dot{V}(\xi_i)\right] \quad (6)$$

$$a_4 = \frac{4}{h_i^4}\left[V(\xi_{i-1}) + 4V\left(\xi_{i-1} + \frac{1}{2}h_i\right) + V(\xi_i)\right]$$
$$-\frac{4}{h_i^4}\left[\dot{V}(\xi_{i-1})(2\xi_{i-1} + 3\xi_i) + 10\dot{V}\left(\xi_{i-1} + \frac{1}{2}h_i\right)(\xi_{i-1} + \xi_i) + \dot{V}(\xi_i)(3\xi_{i-1} + 2\xi_i)\right] \quad (7)$$
$$-\frac{24}{h_i^5}\left[V(\xi_{i-1})(2\xi_{i-1} + 3\xi_i) - V(\xi_i)(3\xi_{i-1} + 2\xi_i)\right]$$

The maximum of $|R(t)|$ satisfies

$$|R(t)|_{max} \leq \frac{|V^{(4)}(\eta)|_{max}}{4!}\left|[(t-\xi_{i-1})(t-\xi_i)]^2\right|_{max}$$
$$\leq |5a_5\eta + a_4|_{max}\left(\frac{1}{4}h_i^2\right)^2 \quad (8)$$

Given a tolerance value of the approximation error, denoted as $\varepsilon$, the maximum time step should be

$$\hat{h}_i = \left(\frac{16\varepsilon}{|5a_5\eta + a_4|_{max}}\right)^{1/4} \quad (9)$$

The maximum time step can be iteratively corrected. The termination condition of the iteration is given as follows

$$\frac{\left|(\hat{h}_i)_k - (\hat{h}_i)_{k-1}\right|}{(\hat{h}_i)_{k-1}} \leq \mu \quad (10)$$

where $k$ is the iteration number and $\mu$ is the tolerance ratio of $\hat{h}_i$. Note that the tolerance of approximation determines the interpolation accuracy and also affects the computational speed.

After obtaining the maximum time step, cubic Hermite interpolation is implemented to fit $V(t)$ according to Eq. (3). Rise and set times are further obtained using analytical formulas of roots of cubic polynomials. The self-adaptive Hermite interpolation technique is suitable for multiple hump function approximation and root finding. It guarantees the approximation accuracy with the lowest number of interpolation points. Thus the computation cost is reduced to the minimum.

## 2.2. Algorithm design

The first-order derivative of the visibility function is given by

$$\dot{V}(t) = \frac{1}{\|\Delta r\|}\left(\Delta\dot{r}\cdot r_{site}^0 + \Delta r\cdot \dot{r}_{site}^0\right) - \frac{1}{\|\Delta r\|^3}(\Delta r\cdot \Delta\dot{r})\Delta r\cdot r_{site}^0 \quad (11)$$

where $\Delta\dot{r}$ equals the velocity difference between the satellite and the ground target

$$\Delta\dot{r} = v_{sat} - v_{site} \quad (12)$$

All the vectors in Eq. (11) are with respect to the Earth-centered inertial (ECI) frame. The satellite position and velocity can be computed using any analytical orbit propagator, not limited to Keplerian orbits. The position and velocity vectors of the ground target can be obtained using a simple Z-axis rotation model or the IERS Earth orientation model [13].

By applying the mathematical model described in Subsection 2.1, we can derive the detailed algorithm for in-view period computation. As depicted in Fig. 3, the overall algorithm design is summarized as follows:

1) Given the initial position and velocity information of the satellite and ground target, and the time interval $[t_1, t_2]$, let $i = 1$ and $\xi_{i-1} = t_1$.

2) Set the initial value for the current interpolation step. Let $k = 1$. If $i = 1$, $(\hat{h}_i)_{k-1} = 100$ s; else $(\hat{h}_i)_{k-1} = h_{i-1}$.

3) Compute $V(t)$ and $\dot{V}(t)$ at $t = \xi_{i-1}$, $\xi_{i-1} + (\hat{h}_i)_{k-1}/2$, $\xi_{i-1} + (\hat{h}_i)_{k-1}$ using Eqs. (1) and (11). Then obtain the approximate function of $V^{(4)}(t)$ using Eq. (5).

4) Correct $(\hat{h}_i)_{k-1}$ according to Eq. (9) using $V^{(4)}(t)$ and the user-defined approximation error tolerance $\varepsilon$. If Eq.

(10) is satisfied, iteration terminates and $h_i = (\hat{h}_i)_{k-1}$; else $k = k+1$ and go back to 3).

5) Let $\xi_i = \xi_{i-1} + h_i$. Implement cubic Hermite interpolation using Eq. (3) on the subinterval $[\xi_{i-1}, \xi_i]$.

6) Determine whether there are roots on the current subinterval. If so, obtain the cubic polynomial roots using analytical formulas.

7) If $\xi_i < t_2$, $i = i + 1$ and go back to 2); else, the algorithm terminates.

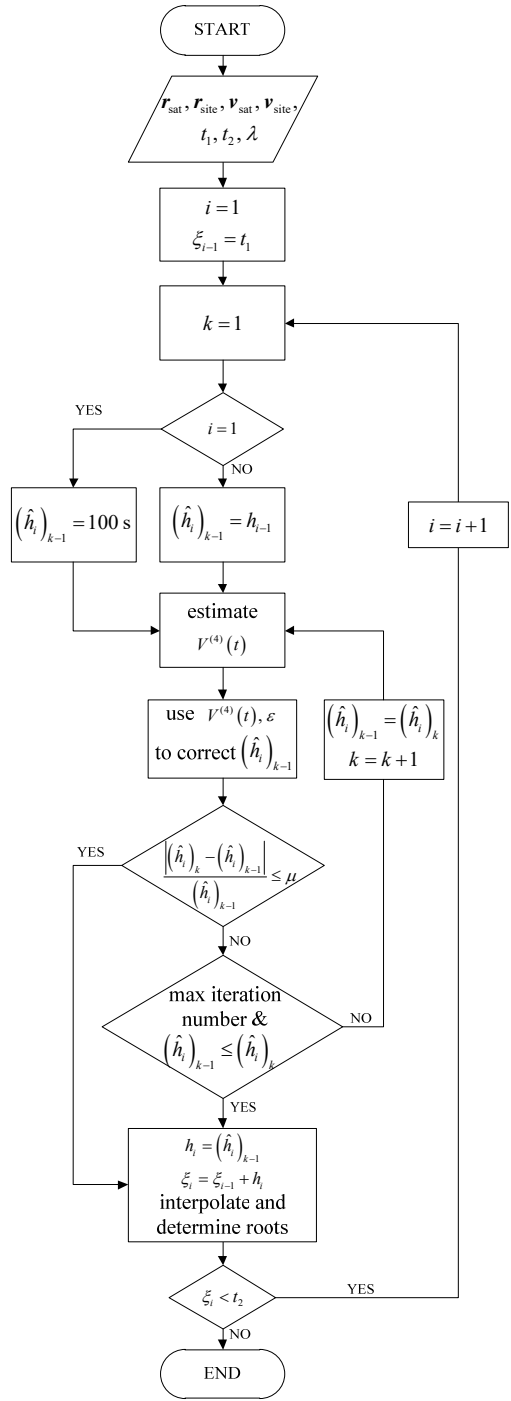

**Figure 3**. Flowchart of the self-adaptive interpolation algorithm for visibility determination.

## 3. Interval shrinking using viewing cone

The satellite could not see the ground target during several continuous orbital periods. As seen in Fig. 2, the satellite never rises above the minimum elevation from 10:00 to 20:00. In order to explain this phenomenon, we define the viewing cone of the ground target (see Fig. 4). The vertex of the infinite cone is located at the ground target. Its axis is along the $r_{site}$ direction, and its half cone angle equals $\pi/2 - \theta_0$. The viewing cone is fixed on the Earth's surface and rotates around the Earth's axis in the inertial space, whereas the satellite trajectory is constrained in a nearly inertially-fixed orbital plane. The satellite sees the ground target only when it moves in the intersecting arc (thick solid line) between the viewing cone and the orbit trajectory.

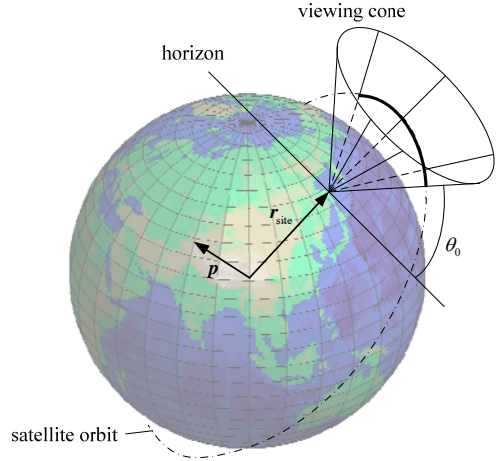

**Figure 4.** Geometric relationship between the ground viewing cone and the satellite's orbit trajectory.

An analytical method has been developed to screen out non-intersecting periods. Consider the situation in which the ground viewing cone is tangent to the orbit trajectory. The tangent point corresponds to the boundary of the intersecting period. Let $q$ denote the vector from the Earth center to the tangent point and $p$ the orbital angular momentum vector ( $p = r_{sat} \times v_{sat}$ ). It can be easily proven that the three vectors $r_{site}$, $p$, and $q$ are in the same plane. Figure 5 shows a cutaway view of the three vectors. Define $\gamma$ as the angle between $p$ and $r_{site}$ at the tangent point. $\gamma$ is related to $\theta_0$ as follows

$$\gamma = \theta_0 + \arcsin \frac{\|r_{site}\| \sin(\pi/2 + \theta_0)}{\|q\|} \tag{13}$$

For a circular orbit, $\|q\|$ always equals the orbital radius. For an elliptical orbit, $\|q\|$ varies with true anomaly of the tangent point. In this study, the maximum orbital radius is used to represent $\|q\|$ for simplicity.

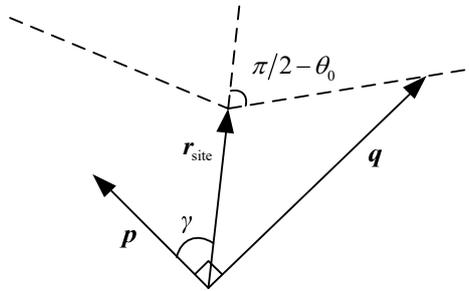

**Figure 5.** Cutaway view of $r_{site}$, $p$, and $q$ at the tangent point.

The intersection criterion can be given as follows

$$\Omega = \frac{\boldsymbol{p} \cdot \boldsymbol{r}_{\text{site}}^0}{\|\boldsymbol{p}\|} \leq \cos\gamma \tag{14}$$

$\Omega$ varies with time and can be further expressed as

$$\Omega = \tilde{p}_x \cos\varphi_0 \cos(\omega_E t + \phi_0) + \tilde{p}_y \cos\varphi_0 \sin(\omega_E t + \phi_0) + \tilde{p}_z \sin\varphi_0 \tag{15}$$

where $\tilde{p}_x$, $\tilde{p}_y$, and $\tilde{p}_z$ are the components of the unit vector of $\boldsymbol{p}$, $\varphi_0$ and $\phi_0$ are the geocentric latitude and longitude of the ground target (in the ECI frame) at the initial epoch, and $\omega_E$ is the angular velocity of the Earth rotation. The in/out bounds of the intersecting periods can be analytically obtained as follows

$$t_{\text{in/out}} = \frac{1}{\omega_E}\left(\arcsin\frac{\cos\gamma - \tilde{p}_z \sin\varphi_0}{\sqrt{\tilde{p}_x^2 + \tilde{p}_y^2}\cos\varphi_0} - \phi_0 - \arctan\frac{\tilde{p}_x}{\tilde{p}_y} + 2\pi m\right), \quad m = 0, 1, 2, \ldots \tag{16}$$

or

$$t_{\text{in/out}} = \frac{1}{\omega_E}\left(\pi - \arcsin\frac{\cos\gamma - \tilde{p}_z \sin\varphi_0}{\sqrt{\tilde{p}_x^2 + \tilde{p}_y^2}\cos\varphi_0} - \phi_0 - \arctan\frac{\tilde{p}_x}{\tilde{p}_y} + 2\pi m\right), \quad m = 0, 1, 2, \ldots \tag{17}$$

With $t_{\text{in/out}}$ we can get rid of the non-intersecting periods to shrink the overall searching interval for visibility determination.

## 4. Numerical results

The performance of the rapid satellite-to-site visibility determination algorithm is first evaluated to demonstrate its effectiveness and is then applied to coverage analysis of satellite constellation.

### 4.1. In-view period determination

We consider a Low-Earth-Orbiting (LEO) satellite at an altitude of 1100 km and a ground target located at (110° E, 25° N). The orbital elements, i.e., semi-major axis, eccentricity, inclination, right ascension of ascending node, argument of perigee, and mean anomaly, at the initial epoch are 7478.14 km, 0.05, 50°, 120°, 25°, and 80°, respectively. An analytical orbit propagator incorporating the $J_2$ and $J_4$ perturbation effects is used to compute the satellite position and velocity. The analytical propagator is developed based on the Simplified General Perturbations 4 (SGP4) theory [14]. The minimum elevation angle is set to 10°. Three cases with different approximation error tolerance values, 0.1, 0.01, and 0.001, are carried out. The tolerance ratio for $\left(\hat{h}_i\right)_k$ iteration is set to 0.1.

The true curves of the visibility function as well as the piecewise cubic lines generated for approximation of the three cases are plotted in Fig. 6. It is seen that the piecewise cubic lines (green) achieve good approximation of the waveforms (blue) of the visibility function. Compared with case 1 and case 2, case 3 has the best fitting accuracy due to the use of the smallest error tolerance. The maximum absolute fitting errors are $9.1 \times 10^{-2}$, $9.8 \times 10^{-3}$, and $9.3 \times 10^{-4}$, respectively, for the three cases. The actual errors are close to the error tolerance values, which reflects the error controllability of the method.

The interval shrinking strategy has been successfully implemented and the whole 24-hour data arc of interest is cut down to about 14 hours. The self-adaptive interpolation algorithm is then applied to the interval $[t_{\text{in}}, t_{\text{out}}]$ with $t_{\text{in}} = 1.18$ h and $t_{\text{out}} = 15.08$ h. Table 1 summarizes the rise and set time results. Also listed are the Percentage Normalized Errors. The Percentage Normalized Error (PNE) is defined by [7]

$$\text{PNE} = \frac{|\text{Predicted satellite rise/set time - Acutal satellite rise/set time}|}{\text{Actual in-view period}} \times 100 \tag{18}$$

where the actual rise and set times are computed using a 5-s trajectory checking method. As seen from Table 1, there are 5 in-view periods in the 24 hours. The length of the in-view period varies from 400 to 1000 s. The maximum PNEs for the three cases are 4.6, 1.7, and 0.9, respectively. Reducing the approximation error tolerance leads to increased accuracy of the rise and set time solutions.

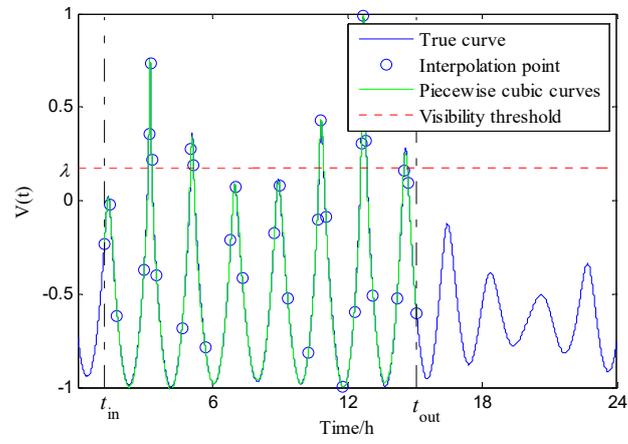

(a) Case 1, ε = 0.1

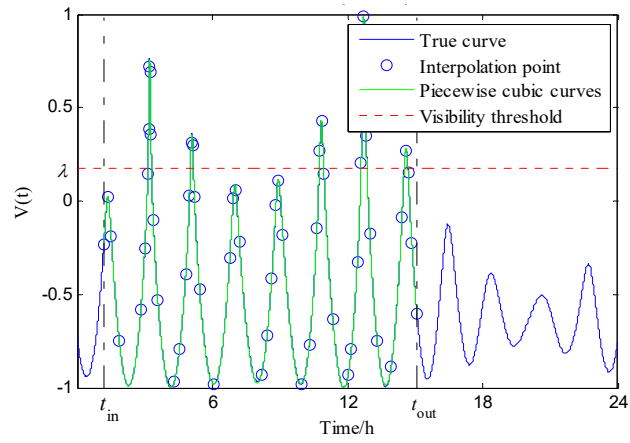

(b) Case 2, ε = 0.01

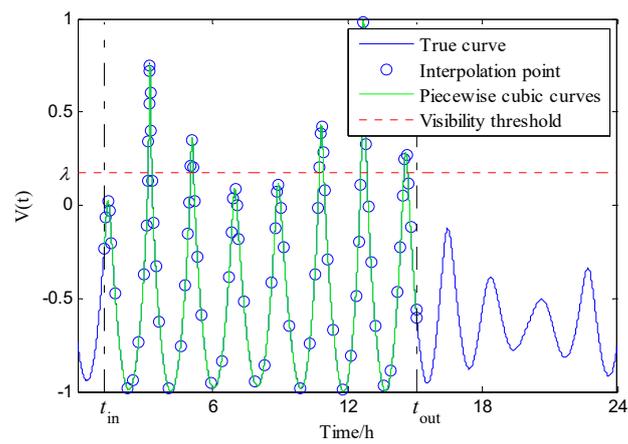

(c) Case 3, ε = 0.001

**Figure 6**. The true curve and piecewise cubic curves for three cases with different approximation error tolerances.

Table 1. One day rise-set summary for the three cases.

| In-view period number | Case 1 ($\varepsilon = 0.1$) | | Case 2 ($\varepsilon = 0.01$) | | Case 3 ($\varepsilon = 0.001$) | |
|---|---|---|---|---|---|---|
| | [$t_{rise}$, $t_{set}$]/s | [$PNE_{rise}$, $PNE_{set}$] | [$t_{rise}$, $t_{set}$]/s | [$PNE_{rise}$, $PNE_{set}$] | [$t_{rise}$, $t_{set}$]/s | [$PNE_{rise}$, $PNE_{set}$] |
| 1 | 11310, 11893 | 0.60, 0.80 | 11302, 11891 | 0.67, 1.16 | 11302, 11893 | 0.67, 0.69 |
| 2 | 18089, 18606 | 3.62, 4.60 | 18098, 18584 | 1.70, 0.16 | 18102, 18581 | 0.88, 0.79 |
| 3 | 38684, 39432 | 0.50, 2.81 | 38679, 39411 | 1.22, 0.08 | 38683, 39408 | 0.60, 0.48 |
| 4 | 45351, 46322 | 0.20, 0.83 | 45349, 46322 | 0.41, 0.81 | 45350, 46326 | 0.39, 0.44 |
| 5 | 52264, 52834 | 0.74, 2.18 | 52250, 52844 | 1.61, 0.52 | 52255, 52843 | 0.71, 0.57 |

The computational speeds of the three cases have been compared with that of the 5-s trajectory checking method. Decreases of 98.16%, 97.14%, and 96.20% in computation time over the brute force method have been achieved. This can also be indicated from the times of calculation of the visibility function. The 5-s trajectory checking method requires 17280 times of calculation, whereas the three cases requires only 278, 462, and 560 times of calculation, counted by multiplying the number of interpolation points with the average number of iterations. In addition, our numerical method is faster than the former methods, i.e., Fourier series representation [8], parabolic blending [9], and APCHI technique [10], which reduce 95%, 95%, and 90% of computation time over the brute force method, respectively.

**4.2. Coverage evaluation**

The rapid visibility determination algorithm is further employed for coverage analysis of the China's BeiDou navigation constellation. By the end of year 2015, the constellation includes 6 satellites in Medium Earth Orbit (MEO), 7 satellites in Inclined Geosynchronous Satellite Orbit (IGSO), and 5 satellites in Geostationary Earth Orbit (GEO). The orbital elements of the 18 satellites are downloaded from the website *https://www.space-track.org*. The analytical orbit propagator incorporating the $J_2$ secular effects and the simple Z-axis Earth rotation model is used. The minimum elevation angle is set to 10°. The approximation error tolerance is set to 0.01. The tolerance ratio for $\left(\hat{h}_i\right)_k$ iteration is set to 0.1.

The case covers a 1-day time span starting from December 21, 2015, 19:00:00.0 (UTC). The area of the Asia-Pacific region referring to [35° E, 160° E] × [60° S, 60° N] has been meshed into grids of 2 × 2 deg (3843 grids). Figure 7 shows the minimum satellite coverage, which measures the minimum number of satellites available simultaneously over the entire coverage interval. It is shown that the BeiDou signals cover the whole China area with a minimum five-fold coverage.

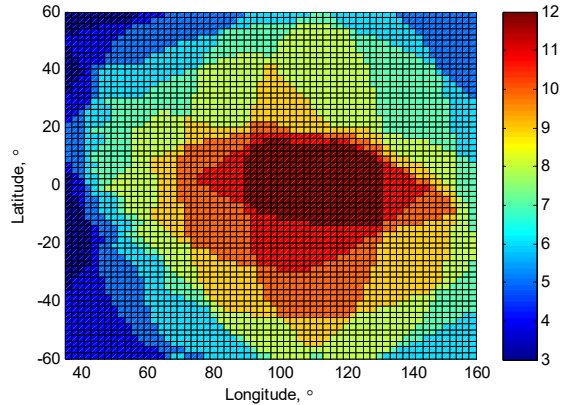

Figure 7. The distribution of minimum satellite coverage of BeiDou signal covering the Asia-Pacific region.

The algorithm is implemented on a personal computer using the Matlab software. The total execution time of visibility determination is 447.6 s. Since Beidou satellites have much longer orbital periods than LEOs. Our method requires only about 10 interpolation points to approximate the visibility function for each Beidou satellite in one day, and a 99.7% decrease in computation time is achieved over the brute force method.

## 5. Conclusion

We have developed a numerical method for rapid determination of satellite-to-site visibility. Compared with the traditional trajectory checking approach, the new method significantly reduces computation time with controllable solution accuracy defined by users. The self-adaptive Hermite interpolation algorithm is applicable for all orbital types and analytical orbit propagators. Actually, more complicated satellite visibility problems, such as imaging opportunity determination considering camera's field of view, can be addressed by our method provided that the mathematical description of visibility function is formulated. Our numerical method has low computational complexity and is attractive for supporting ground-based and onboard autonomous spacecraft operations.

## Acknowledgements

This research is funded in part by Ministry of Science and Technology of China through cooperative agreement No. 2014CB845303.